\documentclass[11pt]{article}
\usepackage{epsfig,dsfont}
\usepackage{geometry}
\begin{document}  
\sffamily

\thispagestyle{empty}
\vspace*{15mm}

\begin{center}

{\huge
Solving the sign problems of the massless lattice\\
\vskip3mm
Schwinger model with a dual formulation}
\vskip25mm
Christof Gattringer, Thomas Kloiber and Vasily Sazonov 
\vskip10mm
University Graz, Institute of Physics\\
Universit\"atsplatz 5, 8010 Graz, Austria 
 
\end{center}
\vskip30mm

\begin{abstract}
We derive an exact representation of the massless Schwinger model on the lattice in terms of dual variables which are 
configurations of loops, dimers and plaquette occupation numbers. 
When expressed with the dual variables the partition sum has only real and positive terms also when a chemical potential or a topological
term are added -- situations where the conventional representation has a complex action problem. The dual representation allows for
Monte Carlo simulations without restrictions on the values of the chemical potential or the vacuum angle.
\end{abstract}

\newpage
\setcounter{page}{1}

\section{Introduction}
Since its initial formulation three decades ago lattice QCD has developed into a reliable quantitative tool 
for studying many low energy phenomena in QCD. However, one important issue where the lattice approach 
has essentially failed so far is its application to QCD at finite density. The reason is that at finite chemical potential 
$\mu$ the fermion determinant becomes complex and cannot be used as a probability weight in a Monte Carlo
simulation. This is known as the ''complex action problem'' or ''sign problem''. 
The complex action problem is not specific for QCD or for theories with fermions, but is a generic feature of many
quantum field theories at finite density both in the lattice and the continuum formulation. Furthermore, not only 
finite chemical potential gives rise to a complex action problem, but also the addition of a topological term. 

For overcoming the complex action problem in lattice simulations several approaches have been discussed,
such as reweighting, various series expansions, the reformulation with alternative variables and simulations based 
on stochastic differential equations. Reviews at the yearly lattice conferences \cite{reviews} give an overview about 
the progress in recent years. 

Probably the most elegant approach to the complex action problem
is to rewrite a lattice field theory in terms of new degrees of freedom, often referred to 
as ''dual variables'', such that in the new representation the partition sum has only real and positive
contributions. The dual variables usually turn out to be loops for matter fields and surfaces for gauge fields. 
The key problem of this approach is that there is no general recipe for finding a dual representation of a given 
theory. In particular non-abelian gauge fields are difficult and so far no convincing approach has emerged. The situation 
is different for theories with abelian gauge fields where for the case of their interaction with bosonic matter 
several interesting results based on dualization \cite{abelian1,scalarqed2} were presented in recent years.

Coupling fermions to the gauge fields brings in additional difficulties with extra signs for the matter loops 
coming from the Grassmann nature of the fermions and from traces over gamma-matrices.
In 2 dimensions, however, the understanding of the traces over gamma-matrices as they appear in dual fermion loops is much more 
advanced \cite{nucu}, such that the Schwinger model, i.e., QED in 2 dimensions, seems to be an interesting candidate for further 
developing the dual approach for theories of gauge fields interacting with fermions.  
Indeed, some results for at least partial dualizations of the lattice Schwinger model were discussed in the literature
\cite{schwinger1, schwinger2, schwinger3}, but no convincing complete solution of the sign problem was presented so far. 

In this article we show that for the lattice Schwinger model with massless staggered fermions the complex action problem can be solved
completely by exactly rewriting the model to a dual representation. We consider both types of complex action problems, coming from finite density 
and the addition of a topological term. The dual variables are closed oriented loops for the fermions, with the gauge fields being
represented by integer valued plaquette occupation numbers. The constraints for these dual degrees of freedom again allow for the 
interpretation of the dual gauge field variables as surfaces in 2 dimensions. We show that in the dual form the partition sum has only real and
positive contributions. Furthermore we discuss the dual formulation of various observables and briefly address possible update strategies.
  
We stress that the dual representation discussed here not only provides an interesting step towards solving complex action problems for theories 
of gauge fields interacting with fermions, but that it will also be useful for addressing some of the open questions concerning the phase structure and 
universality class of the massless lattice Schwinger model with staggered fermions \cite{durr,shailesh}.

\section{Conventional form of the one flavor model with topological term}

We begin with the case of the massless one-flavor model with a topological term. In the conventional representation 
the corresponding partition sum 
is given by
\begin{equation}
\label{eq_conv_part}
Z  \; = \; \int \! \mathcal{D}[U]\mathcal{D}\big[\,\overline{\psi},\psi \,\big]~e^{-\, S_G[U]  \, -i\,\theta \, Q[U] \, - \, S_\psi[U,\overline{\psi},\psi] }  \; .
\end{equation}
The dynamical degrees of freedom are U(1)-valued link variables $U_\nu(n) = \exp(i A_\nu(n)), \, A_\nu(n) \in [-\pi, \pi]$. Here
$n = (n_1,n_2)$ denotes the sites of a 2-dimensional $N_S \times N_T$ lattice and $\nu = 1,2$ runs over the Euclidean space ($\nu = 1$) and 
time ($\nu = 2$) directions. The second set of degrees of freedom are the one-component (we work with staggered fermions)
Grassmann valued fermion variables $\overline{\psi}(n)$ and $\psi(n)$. All boundary conditions are periodic, except 
for the temporal boundary conditions of the fermions which have to be chosen anti-periodic. By $V = N_S N_T$ we denote 
the total number of lattice points.

The measures in the path integral are the product measures of U(1) Haar-measures for the gauge fields and Grassmann measures
for the fermions:
\begin{equation}
\label{measures}
\int \! \mathcal{D}[U]  \; = \; \prod_{n,\nu} \int_{-\pi}^\pi \! \frac{d A_\nu (n)}{2 \pi} \; \; \; , \; \; \;  \; \;
\int \! \mathcal{D}\big[\, \overline{\psi},\psi \, \big] \; = \; \prod_{n} \! \int d\overline{\psi}(n) \, d \psi(n) \; .
\end{equation}
For the gauge action we use the Wilson form (the constant term was dropped for simplicity), 
\begin{equation}
\label{eq_conv_gauge}
S_G[U] \; = \; -\beta \sum_{n} \mbox{Re} \, U_p(n)  \; = \; -\frac{\beta}{2} \sum_{n} \big[ \, U_p(n) \, + \, U_p(n)^{-1} \, \big] \; .
\end{equation}
The plaquette variables $U_p(n) \, = \,U_1(n) \, U_2(n+\hat{1}) \, U_1 (n+\hat{2})^{-1} \, U_2(n)^{-1}$ are the usual
products of the link variables. We remark, that in two dimensions all plaquettes $U_p(n)$ can be labelled by the coordinate $n$
of their lower left corner. For the topological charge $Q[U]$ we use the field theoretical definition 
\begin{equation}
\label{eq_conv_topcharge}
Q[U] \; = \; \frac{1}{i 4 \pi} \sum_n \big[ \, U_p(n) \, - \, U_p(n)^{-1} \, \big] \; .
\end{equation}
We stress that the field theoretical definition (\ref{eq_conv_topcharge})  approaches 
the continuum form $Q = 1/2\pi \int \!d^2x ~F_{12}(x)$ only in the continuum limit. At finite lattice spacing
$Q[U]$ is not an exact integer and $\theta$ is not an angle and consequently requires renormalization. 
In a recent publication \cite{scalarqed2} it was shown non-perturbatively in a lattice simulation of scalar 
QED$_2$ how a suitable continuum limit can be set up, such that $Q[U]$ as defined in (\ref{eq_conv_topcharge})
converges to the continuum form and the $\theta$-dependence of observables becomes $2\pi$-periodic, i.e., 
$\theta$ indeed becomes an angle.  

The action for massless staggered fermions is given by 
\begin{equation}
\label{psiaction}
S_\psi[U,\overline{\psi},\psi] \; = \; \frac{1}{2} \sum_{n, \nu } \Big[ \gamma_\nu(n) \, U_\nu(n) \, \overline{\psi}(n) \, \psi(n +\hat{\nu}) \; - \; 
\gamma_\nu(n) \, U_\nu(n)^{-1} \, \overline{\psi}(n + \hat{\nu}) \, \psi(n) \Big] \; ,
\end{equation}
where the staggered sign function $\gamma_\nu(n)$ is defined as
\begin{equation}
\gamma_1(n) \; = \; 1 \; \; \; , \; \; \; \; \; \gamma_2(n) \; = \; (-1)^{n_1} \; .
\end{equation}
 
Putting things together we arrive at the following form of the partition function
\begin{equation}
\label{partsum}
Z  \; = \; \int \!\! \mathcal{D}[U]\mathcal{D}\big[ \, \overline{\psi},\psi \, \big] \;  \; 
e^{ \, \eta \sum_n U_p(n)} \; \; e^{ \, \overline\eta \sum_n U_p(n)^{-1}} \; 
e^{  - \, S_\psi[U,\overline{\psi},\psi] } \; ,
\end{equation}
where we have introduced the abbreviations
\begin{equation}
\eta \;  = \; \frac{\beta}{2}-\frac{\theta}{4\pi}\; \; \; , \; \; \; \; \; \overline{\eta} \; = \; \frac{\beta}{2}+\frac{\theta}{4\pi} \;.
\end{equation}
It is obvious from (\ref{partsum}) that the conventional representation has a complex action problem for non-zero values of the \(\theta\) angle. 
Then $\eta \neq \overline\eta\,$ and the Boltzmann factor acquires a phase and is therefore not suitable as a probability weight for importance 
sampling. We remark, that for the physically interesting region where we want to construct the continuum limit ($\beta \rightarrow 
\infty$) both parameters $\eta$ and $\overline{\eta}$ are positive. We will assume this positivity from now on, i.e., we restrict ourselves to
$\beta > 1/2$ with $\theta \in [-\pi,\pi]$.

For later use we redefine the link variables $U_\nu(n)$ by absorbing the staggered sign function $\gamma_\nu(n)$:
\begin{equation}
\label{linktrafo}
U_\nu^\prime(n) \; = \; \gamma_\nu(n) \, U_\nu(n) \; .
\end{equation}
This transformation obviously removes the staggered signs from the fermion action (\ref{psiaction}). The plaquette simply changes sign,
\begin{equation}
U_p^\prime(n) \, = \,
U_1^\prime(n)  U_2^\prime (n+\hat{1})  U_p^\prime (n+\hat{2})^{-1}  U_2^\prime(n)^{-1} \, = \; 
- \, U_1(n)  U_2(n+\hat{1}) U_1 (n+\hat{2})^{-1}  U_2(n)^{-1} \, = \, - \, U_p(n) \, .
\end{equation}
Due to the fact that the values $\pm 1$ of the staggered signs are elements of the gauge group U(1)
the path integral measure for the gauge fields remains invariant 
($\int \! \mathcal{D}[U^\prime] = \int \! \mathcal{D}[U]$) and we 
obtain for the partition sum (for notational convenience we drop the primes on the transformed link variables)
\begin{equation}
Z \; = \;  \int \! \mathcal{D}[U] \; e^{ \, - \eta \sum_n U_p(n)} \;  e^{ \, - \overline\eta \sum_n U_p(n)^{-1}} \; Z_\psi[U] \; = \; 
 \int \! \mathcal{D}[U] \; \prod_{n} e^{ \, - \eta \, U_p(n)} \;  e^{ \, - \overline\eta \, U_p(n)^{-1}} \; Z_\psi[U] \; ,
\label{znew} 
\end{equation}
where we have defined the fermionic partition sum $Z_\psi[U]$ in a background gauge field $U$ as 
\begin{equation}
Z_\psi[U] \; = \;  \int \! \mathcal{D}\big[ \, \overline{\psi},\psi \, \big] \; 
e^{ \, - \frac{1}{2} \sum_{n, \nu } [ \, U_\nu(n) \, \overline{\psi}(n) \, \psi(n +\hat{\nu}) \; - \; 
U_\nu(n)^{-1} \, \overline{\psi}(n + \hat{\nu}) \, \psi(n) ] } \; .
\label{zf}
\end{equation}
With the transformation (\ref{linktrafo}) we have removed the staggered signs and traded them for an extra minus sign 
for the gauge action (and the 
topological charge term). 

We remark that the transformation can be applied in arbitrary numbers of dimensions and for all theories where
$-1$ is in the gauge group (U(1), SU(2N) et cetera). It always has the same effect of removing the staggered signs in the fermion action and 
flipping the sign of the Wilson gauge action.

\section{Integrating over the fermion fields}

The next step is to integrate out the fermions in the fermionic partition sum $Z_\psi[U]$ defined in (\ref{zf}). 
For that purpose the Boltzmann factor in $Z_\psi[U]$ is expanded, 
such that the Grassmann integral can be saturated with the terms from the expansion. 
The integral is saturated when at each site $n$ both Grassmann variables $\psi(n)$ and 
$\overline{\psi}(n)$ appear in the integrand as factors generated from the expansion. 
In other words, the only non-vanishing Grassmann integral  is
\begin{equation}
\int \! \mathcal{D}\big[ \, \overline{\psi},\psi \, \big] \prod_n \psi(n) \overline{\psi}(n) \; = \; 
\int \! \prod_n d\overline{\psi}(n)  d\psi(n)  \,  \psi(n) \overline{\psi}(n) \; = \; 1 \; .
\label{masterintegral}
\end{equation}
We proceed by rewriting the sum in the exponent of the Boltzmann factor of  (\ref{zf}) into a product and expand the individual exponentials
\begin{eqnarray}
Z_F[U] & = &  \!\int \! \mathcal{D} \big[\, \overline{\psi},\psi \, \big] \; 
\prod_{n, \nu } \; e^{ \, - \frac{1}{2} \, U_\nu(n) \, \overline{\psi}(n) \, \psi(n +\hat{\nu})} \; 
e^{\, \frac{1}{2} \, U_\nu(n)^{-1} \, \overline{\psi}(n + \hat{\nu}) \, \psi(n) }
 \\
& = & \!\int \! \mathcal{D}\big[ \, \overline{\psi},\psi \, \big] \; \prod_{n, \nu } \;
\sum_{k_\nu(n) = 0}^1 \!\! \left(\!- \frac{U_\nu(n)}{2} \, \overline{\psi}(n) \, \psi(n\!+\!\hat{\nu})\! \right)^{k_\nu(n)}  \; 
\sum_{\overline{k}_\nu(n) = 0}^1 \!\!\left( \frac{U_\nu(n)^{-1}}{2}  \, \overline{\psi}(n\!+\!\hat{\nu}) \, \psi(n)\! \right)^{\overline{k}_\nu(n)} \; .
\nonumber
\end{eqnarray}
Since the Grassmann variables are nilpotent, the power series for the exponentials terminate after the linear term. Thus the summation indices
$k_\nu(n)$ and $\overline{k}_\nu(n)$ which we use for the forward and backward hopping terms run only over the values $0$ and $1$, and 
we will refer to them as ''activation indices'' for the corresponding terms. The terms that are activated, i.e., those with 
$k_\nu(n)$, $\overline{k}_\nu(n) = 1$, bring as factors the corresponding 
link variables $U_\nu(n)$ for forward hops and $U_\nu(n)^{-1}$ for backward hops, as well as trivial factors of 
$1/2$ (which, as we will see, combine to an overall factor of $(1/2)^V$). In addition all activated forward hops contribute a factor of $-1$.  

The simplest way for saturating the Grassmann integrals on two neighboring sites $n$ and $n+\hat{\nu}$ is to activate the 
forward and the backward hopping terms that connect the two sites, i.e., one sets $k_\nu(n) = \overline{k}_\nu(n) = 1$.
Such a term is referred to as ''dimer'' and the corresponding contribution is
\begin{equation}
- \frac{U_\nu(n)}{2} \frac{U_\nu(n)^{-1}}{2} \int \!  d\overline{\psi}(n\!+\!\hat{\nu}) d\psi(n\!+\!\hat{n})  d\overline{\psi}(n)  d\psi(n) \; 
\overline{\psi}(n) \psi(n\!+\!\hat{\nu}) \overline{\psi}(n\!+\!\hat{\nu}) \psi(n)  \; = \; \frac{1}{4} \; .
\end{equation}
The link variables cancel and the overall minus sign which comes from the forward hop is compensated by a minus sign from the interchanges of 
Grassmann variables when the last factor $\psi(n)$ of the integrand is commutated to the beginning of the integrand in order to have the
Grassmann variables in the canonical ordering of the integral (\ref{masterintegral}). Thus, up to a factor of $1/4$, which we will absorb in
an overall constant for the partition sum $Z_\psi[U]$, a dimer saturates the Grassmann integrals trivially. 

The second type of contributions that saturate the Grassmann integral are oriented closed loops, where the activation indices  
$k_\nu(n)$ and $\overline{k}_\nu(n)$ along the contour of a loop are equal to $1$. Obviously a hopping term in the loop 
provides a $\psi(n)$ and the next hopping term in the loop the corresponding $\overline{\psi}(n)$ to saturate the Grassmann integrals
for all sites in the loop. We obtain an overall minus sign from commuting the last $\psi(n)$ to the very beginning of the integrand. 
Thus each loop comes with an overall minus sign. In addition every forward hop in the loop comes with a minus sign and, since for a closed loop
the number of forward hops is half of the total number of hops in the loop, we can write this sign as $(-1)^{L(l)/2}$, where $L(l)$ denotes the length of
the loop $l$. This counting of the signs from forward hops is also correct for loops that close by winding around the boundary if we 
choose the number of sites in each direction to be a multiple of $4$ (otherwise trivial corrections have to be taken into account). For loops
that close around the compact time direction we have an additional minus sign from 
the anti-periodic temporal boundary conditions of the fermions, but
we will show below that the total temporal winding number of all loops must be even and this sign will turn out to be irrelevant. Furthermore, each 
hop comes with a factor of $1/2$ and the corresponding link variables $U_\nu(n)$ for forward hops and $U_\nu(n)^{-1}$ for backward hops.
Thus the overall contribution of a loop $l$ is:
\begin{equation}
- \; \left( \frac{1}{2} \right)^{L(l)} \, (-1)^{W(l)} \, (-1)^{\frac{1}{2} L(l)}  \prod_{(n,\nu)\in l} U_\nu(n)^{s_\nu(n)} \; .
\end{equation} 
As before $L(l)$ is the length of the loop $l$, i.e., the number of links in the loop 
and $W(l) \in \mathds{Z}$ is the winding number around the compact time direction.
The product is over all links $(n,\nu)$ in the loop $l$, and $s_\nu(n)$ takes into account the orientation the link is run through in the loop, with
$s_\nu(n) = 1$ for forward hopping and $s_\nu(n) = -1$ for backward hopping.

For completely saturating all Grassmann integrals we can now combine loops and dimers such that each site is either run through by
a loop or is the endpoint of a dimer. Note that due to the absence of mass terms, there are no monomer terms and a non-vanishing 
contribution to $Z_\psi[U]$ must saturate the Grassmann integrals at each site with either a dimer or a loop. Thus $Z_\psi[U]$ is a sum over 
the set $\{l,d\}$ of all configurations of loops and dimers:
\begin{equation}
Z_\psi[U] \; = \; \left( \frac{1}{2} \right)^{V} \sum_{\{l,d\}} (-1)^{N_L} \;  (-1)^{\frac{1}{2} \sum_l L(l)}  \; (-1)^{\sum_l W(l)} 
\; \prod_l \! \prod_{(n,\nu)\in l} U_\nu(n)^{s_\nu(n)} \; ,
\label{zfdual}
\end{equation}
where $N_L$ denotes the total number of loops in a configuration. The factors $1/2$ from the hops in the dimers and 
in the loops combine into a trivial overall factor of $(1/2)^V$.

\begin{figure}[t!]
\begin{center}
\includegraphics[width=7cm,type=pdf,ext=.pdf,read=.pdf]{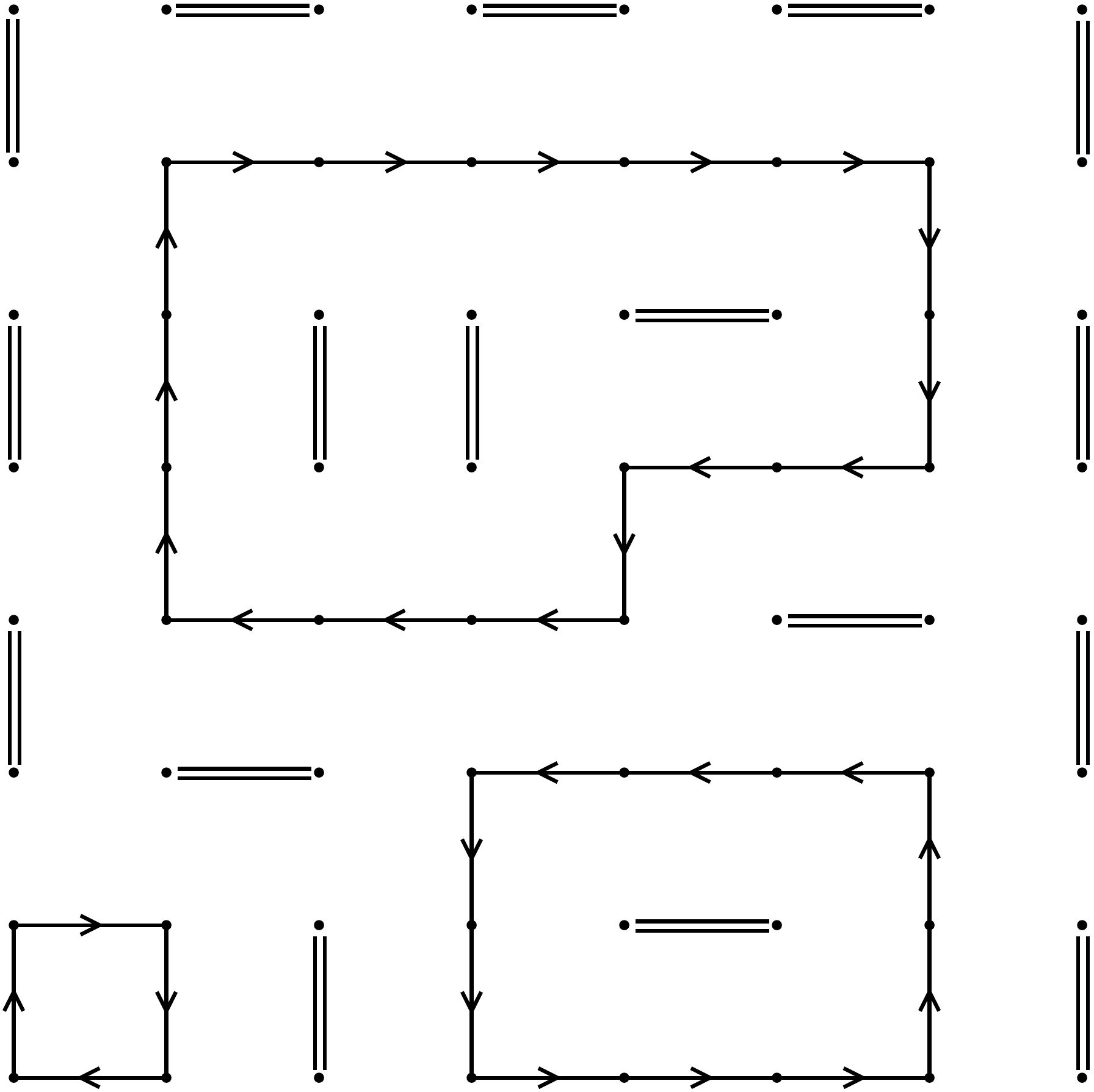} 
\hskip22mm
\includegraphics[width=7cm,type=pdf,ext=.pdf,read=.pdf]{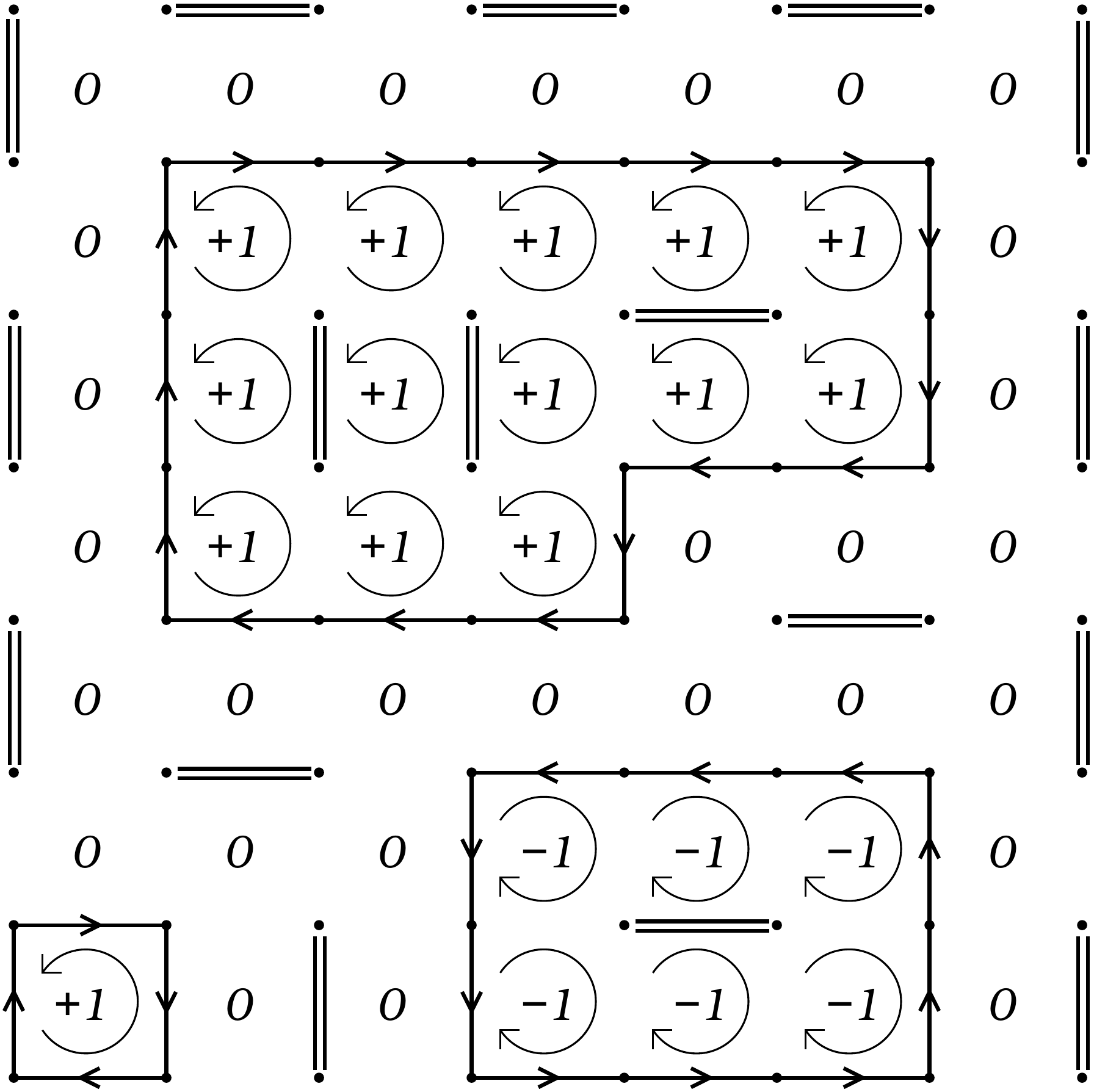}
\end{center}
\caption{Example of an admissible configuration on an $8 \times 8$ lattice. In the lhs.\ plot we show only the fermion contribution, i.e., the filling of 
the lattice with dimers (the double lines on the links) and with oriented loops. In the rhs.\ plot we also show the plaquette occupation numbers 
needed to compensate the link flux introduced by the loops.}
\label{fig1}
\end{figure}

In the lhs.\ plot of  Fig.~\ref{fig1} we show a legitimate configuration of loops and dimers that contributes to the fermionic partition sum 
(\ref{zfdual}): Each site is either run through by an oriented loop or is the endpoint of a dimer. Note that this constraint 
restricts the possible shapes of the loops, and, e.g., a $2 \times 2$ loop is not possible since the single site inside the loop 
cannot be saturated by a dimer. In more than two dimensions or in the presence of monomers from mass terms a $2 \times 2$ 
loop would be possible.  

\section{Integrating over the gauge fields}

After having found the representation of the fermionic partition sum $Z_\psi[U]$ in terms of dimer and loop configurations, we 
now can integrate out the gauge links. The gauge links come from two different sources: Each loop in $Z_\psi[U]$ is dressed with the  
link variables $U_\nu(n)$ sitting on its contour, and from the expansion of the Boltzmann factors  
$e^{ \, - \eta \, U_p(n)}$ and $e^{ \, - \overline\eta \, U_p(n)^{-1}}$  
in (\ref{znew}) we obtain further factors of link variables.  Collecting all contributing factors each link variable $U_\nu(n)$ appears with some power 
$j_\nu(n) \in \mathds{Z}$. Using the well known representation of the Kronecker-Delta, 
$1 / 2 \pi \int_{-\pi}^\pi e^{i \phi j} d \phi  = \delta_{j,0}$ for $j \in \mathds{Z}$, we find
the master formula for the gauge links,
\begin{equation}
\int \! \mathcal{D}[U] \; \prod_{n,\nu} U_\nu(n)^{\, j_\nu(n)} \; = \; \prod_{n,\nu} \delta_{j_\nu(n),0} \; .
\end{equation}
The formula implies that for a non-vanishing contribution to the path integral only terms where all links $U_\nu(n)$
appear with exponent $j_\nu(n) = 0$ are admissible, i.e., all link fluxes cancel. 
Thus every link variable $U_\nu(n)$ that is generated by a loop in $Z_\psi[U]$ has to be compensated by a 
factor $U_\nu(n)^{-1}$ from expanding the gauge action terms. In other words, we need to occupy the plaquettes inside the loops 
by powers of  $U_p(n)$ generated from the expansion of $e^{ \, - \eta \, U_p(n)} \, e^{ \, - \overline\eta \, U_p(n)^{-1}}$. This is illustrated in the 
rhs.\ plot of Fig.~\ref{fig1}, where we indicate the occupied plaquettes by their occupation numbers, as well as by arrows to indicate how 
the link fluxes cancel. The arrows of a plaquette at $n$ are oriented in the mathematically positive (negative) sense for $p(n) > 0$ (\,$p(n) < 0\,)$.
No arrow indicates $p(n) = 0$. The figure also illustrates that for links inside a loop the link flux introduced by adjacent plaquettes cancels.

At this point we now also see that the sign $(-1)^{\sum_l W(l)}$ from loops with non-vanishing temporal winding is irrelevant for admissible 
configurations: In order to saturate all gauge links we need to have an equal number of forward and backward winding loops, since adding 
plaquettes only shifts the unsaturated links of a single winding loop. 
Thus the total number of forward and backward winding loops is always even and from now on 
we can drop the factor $(-1)^{\sum_l W(l)}$.

For a convenient algebraic treatment of the gauge integration 
we use the following representation of the Boltzmann factor for the plaquette $U_p(n)$,
\begin{equation}
e^{ \, - \eta \, U_p(n)} \, e^{ \, - \overline\eta \, U_p(n)^{-1}}  \; = \; \sum_{p(n) \in \mathds{Z}} (-1)^{p(n)} \, 
I_{|p(n)|}\!\left(2 \sqrt{\eta \overline{\eta}} \, \right) \, \left( \sqrt{\frac{\eta}{\overline{\eta}}} \, \right)^{p(n)} \; U_p(n)^{ \,p(n)} \; ,
\label{gaugebessel}
\end{equation}
where $I_{|p(n)|}\!\left(2 \sqrt{\eta \overline{\eta}} \, \right)$ 
denotes the modified Bessel functions, and $p(n) \in \mathds{Z}$ is referred to as the plaquette occupation number for the plaquette
at site $n$. The rather elementary proof of (\ref{gaugebessel}) is given in the appendix. Note that for our restriction to positive
$\eta$ and $\overline{\eta}$ the argument of the Bessel functions is real and positive, implying that also 
the values $I_{|p(n)|}\!\left(2 \sqrt{\eta \overline{\eta}} \, \right)$ are real and positive.

Putting things together, we obtain the following result for the full partition sum,
\begin{equation}
Z \; = \;  \left( \frac{1}{2} \right)^{V} \sum_{\{l,d,p\}} (-1)^{N_L \, + \, N_P \, + \, \frac{1}{2} \sum_l L(l)} 
\; \prod_n I_{|p(n)|}\!\left( 2 \sqrt{\eta \overline{\eta}} \, \right) \, \left( \sqrt{\frac{\eta}{\overline{\eta}}} \, \right)^{p(n)} \; ,
\label{zfinal1}
\end{equation}
where we have introduced the total plaquette occupation number 
\begin{equation}
N_P \; = \; \sum_n p(n). 
\end{equation}
The sum $\sum_{\{l,d,p\}}$ in (\ref{zfinal1}) runs over all admissible configurations of loops, dimers and plaquette occupation numbers. The 
admissible configurations have to obey (see the rhs.\ of Fig.~\ref{fig1} for an example):

\begin{itemize}

\item All sites of the lattice have to be either run through by a loop or be the endpoint of a dimer.

\item At all links the fluxes introduced by the fermion loops and from occupied plaquettes must cancel.

\end{itemize}

Some remarks are in order here: For non-vanishing vacuum angle $\theta \neq 0$ we have $\overline{\eta} \neq \eta$, and the factor 
$\left( \sqrt{\eta / \overline{\eta}} \, \right)^{p(n)}$ in (\ref{zfinal1}) 
gives a different weight to positive and negative plaquette occupation numbers. However,
except for the explicit sign $(-1)^{N_L \, + \, N_P \, + \, \frac{1}{2} \sum_l L(l)}$, the weight factor  
is real and positive, i.e., $\prod_n I_{|p(n)|}\!\left( 2 \sqrt{\eta \overline{\eta}} \, \right) \, 
\left( \sqrt{\frac{\eta}{\overline{\eta}}} \, \right)^{p(n)} \in \mathds{R}_+$. Thus, 
if one can show that the sign $(-1)^{N_L \, + \, N_P \, + \, \frac{1}{2} \sum_l L(l)}$ is positive, 
we have successfully transformed the partition sum to a dual representation where only real and positive 
terms appear. The proof that the sign is indeed always positive for admissible configurations of loops, dimers and plaquette 
occupation numbers will be given in the next section. 

\begin{figure}[t!]
\begin{center}
\includegraphics[width=12.8cm,type=pdf,ext=.pdf,read=.pdf]{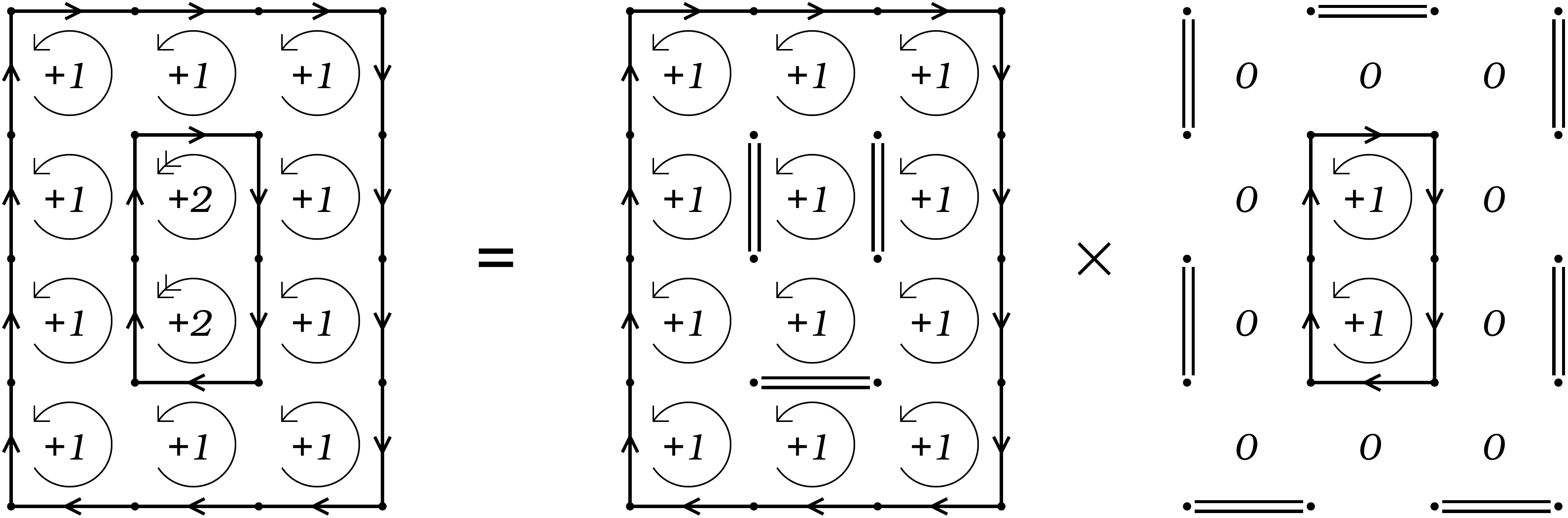} 
\end{center}
\caption{Example for the factorization of a configuration with nested loops. The configuration is decomposed into two factors, 
where in each factor one of the two loops is replaced by a closed chain of dimers.}
\label{fig2}
\end{figure}

Before we come to analyzing the sign we comment on the role of configurations with nested loops. We begin with the simplest case of only 
two loops, an outer one and an inner loop inside the outer loop. If the outer and the inner loop have the same orientation, the plaquette occupation 
numbers inside the inner loop will have $|p(n)| = 2$. If the inner loop has the opposite
orientation of the outer loop, then the plaquette occupation numbers  inside the inner loop will be $p(n) = 0$, i.e., we have a hole. 

The important observation is that concerning the analysis of the sign $(-1)^{N_L \, + \, N_P \, + \, \frac{1}{2} \sum_l L(l)}$, we can decompose the
configuration with the two nested loops into a product of two configurations without nested loops: One factor is the configuration of only the outer 
loop and the inner loop is replaced by a chain of dimers (see Fig.~\ref{fig2}). This replacement of the inner loop by dimers is always possible, since 
a loop has an even number of links and every second link can be replaced by a dimer, which also constitutes an admissible configuration. 
Since the dimers all have weight factor $+1$ this obviously 
does not introduce an extra sign. The second factor is the configuration with only the inner loop. If one multiplies the two factors, then  
also the plaquette occupation numbers take the correct values (see Fig.~\ref{fig2}). 

It is obvious that the factorization trivially generalizes to configurations where we have more than two nested loops. 
The factorization allows one to analyse the sign $(-1)^{N_L \, + \, N_P \, + \, \frac{1}{2} \sum_l L(l)}$ for the simpler configurations
without nested loops, and in the next section we show that this sign is always $+1$. Due to the factorization this then implies that 
\begin{equation}
(-1)^{N_L \, + \, N_P \, + \, \frac{1}{2} \sum_l L(l)} \; = \; +1 \; ,
\label{signformula}
\end{equation}
for all admissible configurations of loops, dimers and plaquette configurations. 

\section{Analysis of the sign factor}
\label{sec_sign}

In this section we show that the sign-formula (\ref{signformula}) holds for all configurations without nested loops, and thus, due to the factorization
property discussed in the end of the previous section, also for all admissible configurations of loops, dimers and plaquettes, including those with
nested loops. 

We begin the proof with showing that for a single loop $l$ of arbitrary shape we have the following relation between its elements:
\begin{equation}
2 \, N_P(l) \; + \; 2 \; = \; 4 \, N_D(l) \; + \; L(l) \; ,
\label{signloop}
\end{equation}
where $N_P(l)$ is the number of plaquettes inside the loop, $N_D(l)$ the number of dimers inside the loop and $L(l)$ the length of the loop,
i.e., the number of links making up the loop. 

\begin{figure}[t!]
\begin{center}
\includegraphics[width=9cm,type=pdf,ext=.pdf,read=.pdf]{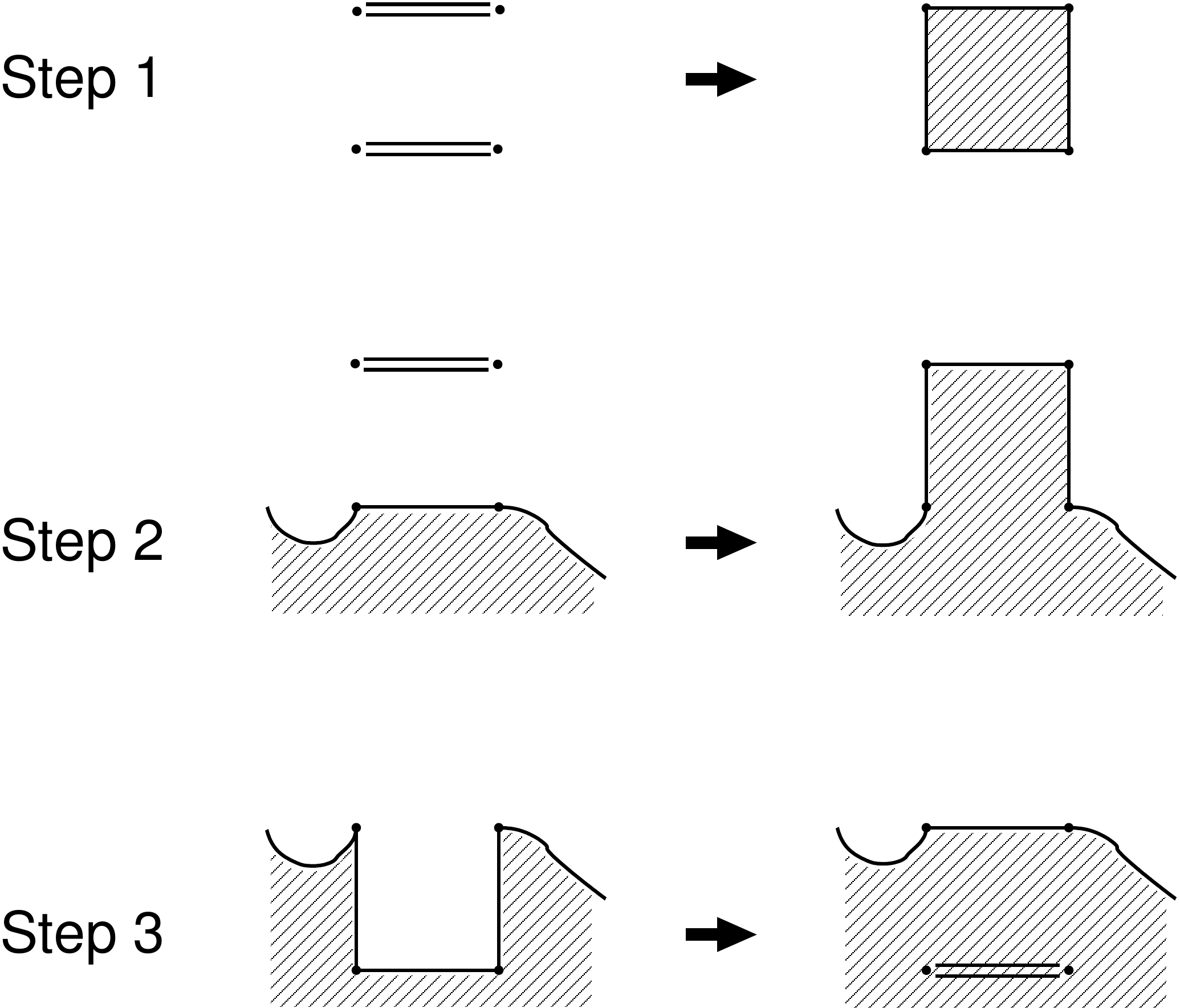} 
\end{center}
\caption{The three steps needed for building up loops: 1) Placing the initial plaquette. 2) Attaching a plaquette at a single link.
3) Attaching a plaquette at three links and placing a dimer. 
For the discussion of (\ref{signloop}) the orientation of the loop is irrelevant and we omit the arrows indicating the
orientation. The loop can be oriented in both, the mathematically positive or the negative sense and the shaded areas indicate 
the corresponding plaquette occupation numbers $+1$ or $-1$.}
\label{fig3}
\end{figure}

\begin{figure}[t!]
\begin{center}
\includegraphics[width=14cm,type=pdf,ext=.pdf,read=.pdf]{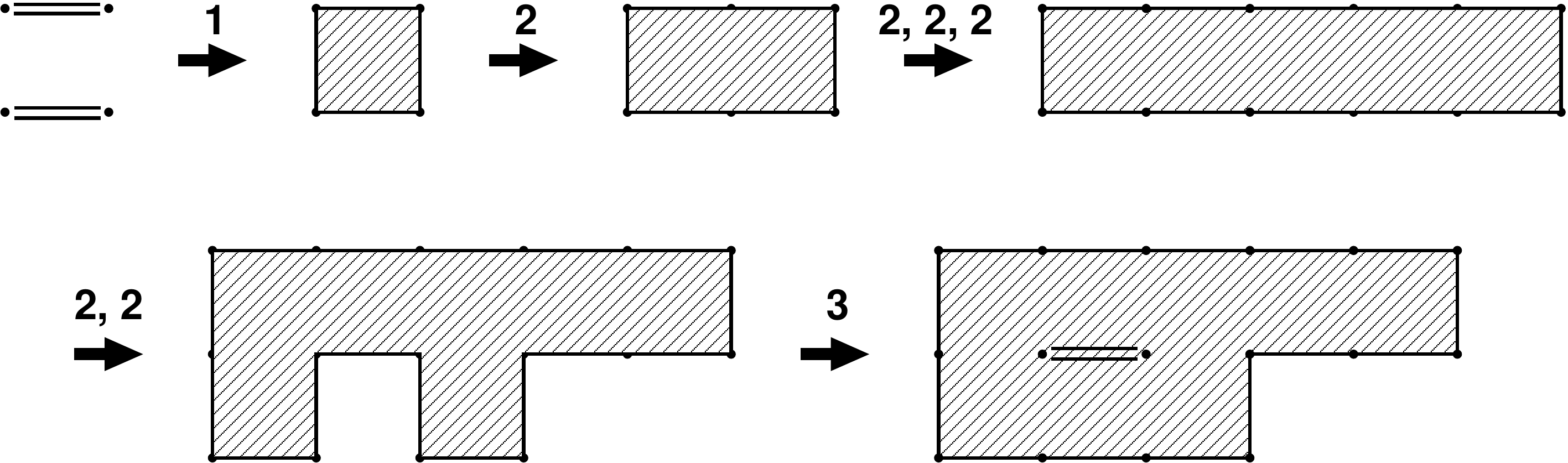} 
\end{center}
\caption{Example of how a loop is built up using the three steps from Fig.~3. Above the arrows we show 
which steps were used to get to the next 
loop. The loop can be oriented in both, the mathematically positive or the negative sense and correspondingly the shaded areas indicate 
plaquette occupation numbers $+1$ or $-1$. For simplicity we here do not show the dimers outside the loop.}
\label{fig4}
\end{figure}

The relation (\ref{signloop}) can be proven by building up an arbitrary loop using a finite number of iterations of three possible steps.
For each step one can analyze the changes $\Delta N_P(l), \Delta N_D(l)$ and $\Delta L(l)$ of the quantities $N_P(l), N_D(l)$ and $L(l)$ which 
characterize a loop. We will see that for each step (\ref{signloop}) remains intact. The three steps are illustrated in Fig.~\ref{fig3}. For the 
analysis of the formula (\ref{signloop}) the orientation of the loop does not play a role and we omit the arrows for the orientation of the loops
in Fig.~\ref{fig3}. The shaded areas indicate plaquettes inside the loop which all have the same plaquette occupation number of either $-1$ for
loops oriented in the mathematically positive sense and $+1$ for loops with negative orientation. Furthermore, as an illustration
in Fig.~\ref{fig4} we give an explicit example for building up a loop using the three steps. 

The three steps are given by:

\begin{enumerate}

\item
Each loop we build starts with placing the first plaquette. This is the simplest loop one can have and its characteristic numbers
\begin{equation}
N_P(l) \; = \; +1 \; , \; \; N_D(l) \; = \; 0 \; , \; \; L(l) \; = \; + 4 \; ,
\end{equation}
obviously obey (\ref{signloop}).

\item
The loop can grow by attaching new plaquettes such that the new plaquette touches only at a single link of the loop. 
This step introduces the changes
\begin{equation}
\Delta N_P(l) \; = \; +1 \; , \; \; \Delta N_D(l) \; = \; 0 \; , \; \; \Delta L(l) \; = \; + 2 \; ,
\end{equation}
which leave (\ref{signloop}) intact.

\item
Finally we can create dimers in the interior of a loop by attaching a new plaquette that touches three links. This step comes with the
changes
\begin{equation}
\Delta N_P(l) \; = \; +1 \; , \; \; \Delta N_D(l) \; = \; 1 \; , \; \; \Delta L(l) \; = \; - 2 \; ,
\end{equation}
which also leave (\ref{signloop}) intact.

\end{enumerate}
After the initial step 1 and a finite number of steps 2 and 3 we arrive at the loop we want to build up. After each step formula (\ref{signloop}) remains 
intact and thus also holds for the final loop. This completes the proof of (\ref{signloop}). 

For establishing the sign formula (\ref{signformula}) we must show that the exponent on the lhs.\ is even, i.e.,
\begin{equation}
N_L \, + \, N_P \, + \, \frac{1}{2} \sum_l L(l) \; = \;  \mbox{even} \; .
\label{signformula2}
\end{equation}
This can be done by summing the formula (\ref{signloop}) which describes the situation for a single loop $l$ over all loops in 
a configuration. We find
\begin{equation}
\sum_l N_P(l) \; + \; \sum_l 1 \; = \; 2 \, \sum_l N_D(l) \; + \; \frac{1}{2} \sum_l L(l) \; ,
\end{equation}
where we have divided (\ref{signloop}) by a factor of 2 on both sides before summing over $l$. The first sum gives the total 
plaquette occupation number $N_P$ (invoke the factorization of configurations). The second sum is simply the number of loops $N_L$.
Rearranging the terms a little one finds
\begin{equation}
N_P \, + \, N_L \, + \, \frac{1}{2} \sum_l L(l) \; = \; 2 \, \sum_l N_D(l) \, + \, \sum_l L(l) \; = \;  \mbox{even} \; ,
\end{equation}
where in the last step we have used that $\sum_l L(l)$ is even since all closed loops can only contain an even number of links. Thus 
 (\ref{signformula2}) holds and the proof of (\ref{signformula}) is complete.

\section{Discussion of the dual representation }
\label{Zdiscussion}

In the previous section we have shown that the sign factors are trivial. Thus we have established the final form of the partition sum,
\begin{equation}
Z \; = \;  \left( \frac{1}{2} \right)^{V} \sum_{\{l,d,p\}}  
\prod_n I_{|p(n)|}\!\left( 2 \sqrt{\eta \overline{\eta}} \, \right) \, \left( \sqrt{\frac{\eta}{\overline{\eta}}} \, \right)^{p(n)} \; .
\label{zfinal2}
\end{equation}
The sum $\sum_{\{l,d,p\}}$ in (\ref{zfinal2}) runs over all admissible configurations of the dual variables,
loops, dimers and plaquette occupation numbers. For each configuration $(l,d,p)$ the weight factors are real and positive, 
and we have completely solved the complex action problem introduced by the topological term (as long as we choose $\eta$ and 
$\overline{\eta}$ positive, i.e., as long as $\beta > 1/2$). 

It is interesting to discuss how the vacuum angle $\theta$ enters the weight factors in the dual representation. For $\theta > 0$ we have 
$\eta < \overline{\eta}$ and thus $\sqrt{\eta/\overline{\eta}} < 1$. From (\ref{zfinal2}) it is obvious that this implies that for $\theta > 0$,
negative plaquette occupation numbers $p(n)$ have a larger weight 
$I_{|p(n)|}\!\left( 2 \sqrt{\eta \overline{\eta}} \, \right) (\sqrt{\eta/\overline{\eta}})^{p(n)}$ 
and thus are favored. In turn this 
implies a larger weight for fermion loops with mathematically positive orientation. However, there are also pure gauge configurations that are 
favored: In the dual representation these are sheets with constant $p(n) < 0$ for all plaquettes, which in the conventional representation 
correspond to configurations with constant field strength, i.e., constant electric field.

Simple bulk observables can now be obtained by taking derivatives of $\ln Z$ with respect to the parameters $\beta$ and $\theta$. 
Evaluating these derivatives also for the dual representation, the dual forms of the topological charge, the topological susceptibility, 
the plaquette expectation value and the plaquette susceptibility can be obtained. 

It is also possible to find the dual representations of various gauge invariant n-point functions. The simplest case are n-point functions of
plaquettes, where one simply can introduce couplings $\eta(n)$ and $\overline{\eta}(n)$ that depend on the site index $n$ and act as sources.
After taking the corresponding derivatives of $Z$ or $\ln Z$,
one sets all $\eta(n)$ and $\overline{\eta}(n)$ to the desired values $\eta$ and $\overline{\eta}$
and so obtains the plaquette n-point functions in the dual representation. 

In a similar way one can construct n-point functions of fermionic currents by multiplying all link variables in the fermion action 
(\ref{psiaction}) by sources $V_\nu(n)$ and $\overline{V}_\nu(n)$ attached to the links for forward and backward hops. 
When integrating out the fermions, the emerging 
loops and dimers are dressed with these sources, while all other steps of the mapping to the dual variables remain unchanged. 
For obtaining the fermionic n-point functions one then evaluates the corresponding derivatives of the dual partition function 
with respect to the sources and then sets all $V_\nu(n) = \overline{V}_\nu(n)= 1$. 

We stress at this point, that the specific real and positive dualization of massless lattice QED presented here is possible only in two dimensions.
From a technical point of view this can be seen from the fact that in more than two dimensions one can find simple configurations which
give negative contributions, e.g., a configuration with a single $2\times2$ loop and the rest of the lattice filled with dimers. From a 
more physical point of view, the successful real and positive dualization of the lattice model reflects the fact that 
2-dimensional fermions can be bosonized also in the continuum, i.e., they can be completely represented in terms of bosonic variables.

\section{Two flavor model with chemical potential}

In the conventional representation the partition sum for the two-flavor model with chemical potential is given by
\begin{equation}
\label{eq_conv_part_2_f}
Z  \; = \; \int \! \mathcal{D}[U]\mathcal{D}[\, \overline{\psi},\psi]\mathcal{D}[\overline{\chi},\chi] 
e^{-\, S_G[U]  \, -i\,\theta \, Q[U] \, - \, S_{\psi}[U,\overline{\psi},\psi]
- \, S_{\chi}[U,\overline{\chi},\chi] }  \; ,
\end{equation}
where $\psi$ and $\chi$ are positively and negatively charged fields 
respectively. The measure $\mathcal{D}[\overline{\chi},\chi]$ for the second flavor $\chi$ 
is a product over Grassmann integrals like
$\mathcal{D}[\,\overline{\psi},\psi]$. The fermionic actions for the two flavors now contain 
chemical potentials $\mu_\psi$ and $\mu_\chi$:
\begin{equation}
\label{chimuaction}
S_{\psi}[U,\overline{\psi},\psi] \; = \; \frac{1}{2} \sum_{n, \nu } \Big[ \gamma_\nu(n) \, e^{\mu_\psi \delta_{\nu,2}} \, U_\nu(n) \, 
\overline{\psi}(n) \, \psi(n \!+\!\hat{\nu}) \; - \; 
\gamma_\nu(n) \, e^{-\mu_\psi \delta_{\nu,2}}\, U_\nu(n)^{-1} \, \overline{\psi}(n \!+ \! \hat{\nu}) \, \psi(n) \Big] ,
\end{equation}
\begin{equation}
\label{chimuaction}
S_{\chi}[U,\overline{\chi},\chi] \; = \; \frac{1}{2} \sum_{n, \nu } \Big[ \gamma_\nu(n) \, 
e^{\mu_\chi \delta_{\nu,2}} \, U_\nu(n)^{-1} \, \overline{\chi}(n) \, \chi(n \!+\!\hat{\nu}) \; - \; 
\gamma_\nu(n) \, e^{-\mu_\chi \delta_{\nu,2}}\, U_\nu(n) \, \overline{\chi}(n \!+ \!\hat{\nu}) \, \chi(n) \Big] .
\end{equation}
We stress that for the second flavor $\chi$ we use the complex conjugate link variables $U_\nu(n)^{\, *}  = U_\nu(n)^{\, -1}$, since
$\chi$ has opposite charge. This is necessary since a theory with U(1) charges at finite density requires overall electric
neutrality due to Gauss' law. 

We remark that for the two-flavor case considered here physics will only depend on the sum of the two chemical potentials $\mu_\psi + \mu_\chi$
due to Gauss' law. However, for more than two flavors (keeping overall electric neutrality) physics will depend also on non-trival combinations of 
the individual chemical potentials (see, e.g., \cite{narayanan}). The generalization of our dualization to more than 
two flavors is trivial, and with this possible generalization in mind we find it instructive to explicitly show the dependence on the individual chemical
potentials in the subsequent derivation of the dualization.
 
The dual representation of the two flavor model is obtained in the same way as in the one-flavor case. 
The first step is to again apply the transformation (\ref{linktrafo}) and subsequently to 
integrate out the fermions. This leads to the loop representations of the fermionic partition sums $Z_\psi[U]$ and $Z_\chi[U]$ 
for the two flavors. The chemical potential terms give rise to extra factors of $e^{\mu_\psi}$ ($e^{\mu_\chi}$) for all forward 
temporal hops of a fermion and factors of  $e^{-\mu_\psi}$ ($e^{-\mu_\chi}$) for backward temporal hops. These factors cancel for
dimers and for loops that close trivially and thus only loops $l$ with a non-trivial winding number $W(l)$ around compact time depend on
the chemical potential via a factor of $e^{\mu_\psi N_T W(l)}$ ($e^{\mu_\chi N_T W(l)}$), where $N_T$ is the temporal lattice extent.  
Thus for the field $\psi$ the dual representation $Z_\psi[U]$ of the fermionic partition sum has the form of
(\ref{zfdual}) with an additional factor $e^{\mu_\psi N_T W(l)}$ and otherwise 
remains unchanged. For the field $\chi$ an additional factor $e^{\mu_\chi N_T W(l)}$ appears in $Z_\chi[U]$, and furthermore all
link variables $U_\nu(n)$ are replaced by $U_\nu(n)^{\, -1}$ due to the opposite charge of $\chi$.

In a second step the product $Z_\psi[U] \, Z_\chi[U] e^{-\,\eta \sum U_p \,  - \, \overline{\eta} \sum U_p^{-1}}$ 
of the two fermion partition sums and the Boltzmann factor for the 
gauge action is integrated over the gauge fields. Again the Boltzmann factors for the individual plaquettes are expanded using 
(\ref{gaugebessel}). Only those terms survive where all link variables from the fermion loops are compensated with links from the plaquettes,
where for the flavor $\chi$ all link variables along the loop are complex conjugate.
The important observation is that one can saturate the links for the loops from the two flavors separately with the plaquettes as needed,
and thus the result for the positivity of the overall sign proven in Section~\ref{sec_sign} can be taken over to show that all configurations
are real and positive also for the two-flavor case. To determine the correct weight for dual gauge field variables one simply
adds the plaquette occupation numbers needed for saturating the link variables for both flavors, and this sum gives the
dual plaquette variable $p(n)$ used for the order of the modified Bessel function $I_{|p(n)|}$ in the weight factor.

The dual form of the partition function of the two flavor case is then given by
\begin{equation}
Z \; = \;  \left( \frac{1}{2} \right)^{2\,V} \sum_{\{l,d,\overline{l},\overline{d},p\}} e^{\mu_\psi \, N_T W(l)} \, e^{\mu_\chi \, N_T W(\overline{l})} 
 \, \prod_n I_{|p(n)|}\!\left( 2\,\sqrt{\eta \overline{\eta}} \, \right) \; \left( \sqrt{\frac{\eta}{\overline{\eta}}} \, \right)^{p(n)} \; .
\label{z2flavorfinal}
\end{equation}
The sum runs over all admissible configurations of the loops and dimers for the $\psi$-field represented as before by $l$ and $d$, 
over all admissible configurations of the loops and dimers for the $\chi$-field represented by $\overline{l}$ and $\overline{d}$, and the 
corresponding admissible configurations of the plaquette occupation numbers $p$. For both flavors, i.e., for both, the 
$l,d$ and the $\overline{l}, \overline{d}$ variables, each site has to be either the endpoint of a dimer or run through by a loop.
For each link the combined flux of the loops of both flavors has to be compensated by activated plaquettes, where the flux from 
the loops $\overline{l}$ that represent the oppositely charged field $\chi$ is counted with a negative sign. 

It is obvious, that also for arbitrary values of the chemical potentials $\mu_\psi$ and $\mu_\chi$ the weights in the 
partition sum (\ref{z2flavorfinal}) are real and positive, and that in the formulation with the dual variables the sign problem is
gone. As discussed for the case of the one-flavor model in Section~\ref{Zdiscussion}, also in the two-flavor case the simplest observables
are obtained as derivatives of $\ln Z$ with respect to the parameters. In particular, here we can also consider the derivatives with respect to 
$\mu_\psi N_T$  and $\mu_\chi N_T$, which give rise to the corresponding particle numbers for the two flavors. The form of the dependence
on the chemical potential in (\ref{z2flavorfinal}) makes clear the interpretation of these particle numbers in terms of the dual variables: They are 
simply given by the temporal net winding numbers $W(l)$ and $W(\overline{l})$. As for the one-flavor case, n-point functions can be obtained 
by coupling local sources, which after taking the corresponding derivatives are set to $1$.

\section{Possible update strategies}

We conclude our presentation of the dual representation of the massless lattice Schwinger model with discussing possible strategies
of a Monte Carlo update. Since in the dual formulation configurations of the variables $l, d$ and $p$ (and for the two-flavor case also 
$\overline{l}$ and $\overline{d}$) have to obey the admissibility conditions, it is not a-priori clear how to propose changes
of a configuration, such that all constraints remain intact. In addition, the proposed changes have to be complete, such that the update is
ergodic, i.e., all configurations can be reached with non-zero probability in a finite number of steps. We do not intend to discuss all technical
details of possible algorithms or their performance (both these issues will be addressed in a forthcoming publication), but we 
sketch the strategies for a dual Monte Carlo simulation such that it is clear that the dual representation is indeed suitable for
simulations without a sign problem.

We begin with noting that the dimers and the elements of the loops come with a Boltzmann factor of 1. For the dimers this implies, 
that we can shift them around for free, and only have to make sure that we modify them such that all fermionic constraints remain intact, 
i.e., all sites are either run through by a loop or are endpoints of dimers (for the two-flavor case this is required for loops and dimers of both flavors). 
For the loops the update is a little more involved, since changing a loop implies also a change of some plaquette occupation numbers $p(n)$
because the flux along the contour of a new piece of loop has to be compensated by suitably occupied plaquettes.  Thus for changing a
loop a Metropolis acceptance step will be necessary, while a change that affects only dimers will always be accepted. Consequently the 
Monte Carlo strategy which we adopt here will consist of alternating updates that affect only dimers with local 
updates that change loops (and might also remove or insert a dimer).  In the case of the two-flavor model,
one has to independently do the dimer-only and the loop updates for both flavors.  
The starting configuration can be chosen as a configuration where the 
whole lattice is filled with dimers. 

We begin with discussing updates where the loops are modified. These updates are essentially patterned according to the three steps
that were used in Section~\ref{sec_sign} to show the positivity of the sign factors and which are illustrated in  Fig.~\ref{fig3}. The difference
for using them in an update is that they can be applied in both directions, i.e., a loop can not only become larger (as in the positivity proof), but
can also shrink or even vanish completely. The steps used in an update of the loops are:

\begin{itemize}

\item 
A plaquette where two sides are occupied by two dimers is replaced by a loop and the corresponding plaquette variable $p(n)$ is changed 
by $\pm 1$ (note that the plaquette under consideration can be nested inside an outer loop such that already before the change we 
have $p(n) \neq 0$. This corresponds to Step 1 of Fig.~\ref{fig3}, and of course can also be run through in both directions, i.e., we can remove
an elementary loop around a single plaquette and replace it by two dimers.

\item 
A plaquette is attached to a loop along a single link and the dimer opposite that link is removed. The plaquette variable $p(n)$ 
is changed by $\pm 1$ depending on the orientation of the loop. This change corresponds to Step 2 in 
Fig.~\ref{fig3} and also can be reverted, i.e., a single-plaquette-detour can be removed from a loop and a dimer is placed. 

\item 
A plaquette is attached to a loop such that it fills a one-plaquette gap in the loop. Three line elements of the loop are replaced by one line
element and a dimer inside of the loop. The plaquette variable $p(n)$ 
is changed by $\pm 1$ depending on the orientation of the loop. This change corresponds to Step 3 in 
Fig.~\ref{fig3} and also can be reverted, i.e., a plaquette and a dimer inside a loop can be removed together.

\item
Finally we need a step that is not depicted in Fig.~\ref{fig3}: We can join two loops of the same orientation that run through opposite links
of a plaquette, by deleting the two loop elements on the two links and inserting them on the two other links of the plaquette. Also here the 
plaquette variable $p(n)$ is changed by $\pm 1$ depending on the orientation of the two loops. The corresponding inverse step cuts a 
loop that has a narrow section with only one plaquette into two loops.

\item
For the system with chemical potential we need another step to populate the different particle number sectors and to explore 
the dependence on the chemical potential. This can, e.g., be done \cite{abelian1} 
by offering double lines of matter flux that wind around the compact time 
direction.

\end{itemize}

All changes are accepted or rejected in a Metropolis step where the acceptance probability is computed from the changing weight 
factors $I_{|p(n)|}$
of the dual gauge variables $p(n)$ that are altered by $\pm 1$. It is easy to see that the steps give rise to an ergodic algorithm for the loops.

As outlined above, the second element of our Monte Carlo strategy is an update of only dimers while keeping the loops fixed in that step.
The dimers have to be placed such that all sites that are not run through by a loop are endpoints of a dimer. As long as this constraint is
obeyed the dimers can be placed randomly since they come with a Boltzmann weight of 1. We thus refer to our update of only the dimers as 
"random dimer insertion". For obeying all constraints the strategy we follow is to first identify the 
dimers where we have no freedom for changing them. An example is given by the dimer inside the last loop of Fig.~\ref{fig4}. These uniquely
determined dimers are frozen and will not be changed. For all sites where the dimers are not determined uniquely we delete the dimers attached 
to that site. Then we start to place dimers randomly connecting empty sites. After each such random placement we again determine whether 
new dimers need to be inserted that are already uniquely determined by the newly placed random dimers. This procedure is iterated until all
dimers are placed and the constraints are obeyed for all sites.

We stress that only the combination of the loop update and the random dimer insertion gives rise to an ergodic algorithm.
Finally, to speed up decorrelation, it is possible to augment the loop and random dimer insertion updates 
with pure gauge updates, where all plaquette variables are changed by 
$\Delta \in \{-1,+1\}$, thus inserting a sheet of constant field strength which otherwise could only be generated by growing a large loop all the 
way around both, the spatial and the temporal boundary. 
Concluding the section about possible algorithms for the dual formulation we remark again, that this discussion 
is only meant to briefly sketch the dual update strategies and a more detailed numerical  analysis is in preparation.

\section{Concluding remarks}

In this paper we have derived a real and positive dual representation of the massless lattice Schwinger model with staggered fermions. 
The dualization solves both sign problems, the one coming from chemical potential and the one introduced by the topological term. 
The dual degrees of freedom turn out to be oriented loops for the fermions, filled with integer valued plaquette occupation numbers
for the gauge fields, which can be interpreted as 2-dimensional surfaces bounded by the fermion loops. All terms in the partition 
sum come with real and positive weight factors, and we have sketched how a simulation in the dual representation can be implemented.
To our knowledge this is the first example of a complete real and positive dualization of a lattice gauge theory with fermions at arbitrary values of the 
chemical potential, of the vacuum angle and of the gauge coupling (up to the trivial restriction $\beta > 1/2$).

The dualization we present is for the case of massless fermions. This is an interesting fact, since so far most attempts for at least a partial
dual representation were starting from a model at infinite fermion mass, i.e., the quenched case, and the dual fermion loops
were obtained from hopping expansion of the fermion determinant, i.e., an expansion in large quark masses. For our dual representation
of the massless Schwinger model it would be interesting to study an expansion of the dual formulation in small mass. 
For non-zero mass one finds negative contributions to the dual partition sum, but for sufficiently small mass this sign problem can be expected 
to be mild and it could be possible to analyze the approach to the chiral limit in a dual representation. 

Another important lesson from dualizing this model theory is the fact that for proving the positivity of all weight factors the inclusion 
of the gauge field dynamics is an essential step. Naively one could have imagined that for the case of finite 
chemical potential, where the complex action problem comes solely from the fermions, one could try to find a positive
representation for the fermionic partition sum alone. However, the construction presented here shows that the gauge field
contributes exactly the necessary signs to make every term in the dual partition sum positive. It can be expected, 
that also for other theories of gauge fields interacting with relativistic fermions a successful dualization strategy will have to 
consider the fermions and the gauge fields on equal footing. 

\vskip5mm
\noindent
{\bf Acknowledgments:} 
Thomas Kloiber is supported by the Austrian Science Fund, 
FWF, DK {\sl Hadrons in Vacuum, Nuclei, and Stars} (FWF DK W1203-N16) and Vasily Sazonov
by the Austrian Science Fund FWF Grant.\ Nr.\ I 1452-N27.
Furthermore this work is partly supported by DFG TR55, {\sl ''Hadron Properties from Lattice QCD''}.

\section*{Appendix}

In this appendix we prove the representation (\ref{gaugebessel}) of the Boltzmann factor for a single plaquette variable.  
For simplicity we here denote the plaquette variable as $U$ with $U \in$ U(1). The formula we want to prove reads
\begin{equation}
e^{ \, - \eta \, U} \, e^{ \, - \overline\eta \, U^{-1}}  \; = \; \sum_{p \in \mathds{Z}} (-1)^{p} \, 
I_{|p|}\!\left(2 \sqrt{\eta \overline{\eta}} \, \right) \, \left( \sqrt{\frac{\eta}{\overline{\eta}}} \, \right)^{p} \; U^{ \,p} \; ,
\label{gaugebessela}
\end{equation}
where $I_{|p|}\!\left( 2 \sqrt{\eta \overline{\eta}} \, \right)$ 
denotes the modified Bessel functions. The first step in the proof is to expand the two exponential functions on the lhs.\ of (\ref{gaugebessela})
into their power series,
\begin{equation}
e^{ \, - \eta \, U} \, e^{ \, - \overline\eta \, U^{-1}} \; = \;  \sum_{k=0}^\infty \frac{(- \eta \, U)^{k}}{k!}
\sum_{\overline{k}=0}^\infty \frac{(-\overline{\eta} \, U^{-1})^{\overline{k}}}{\overline{k}!}
\; = \; 
\sum_{k=0}^\infty \sum_{\overline{k}=0}^\infty \frac{(-1)^{k + \overline{k}}}{k! \, \overline{k}!}
\; \eta^k \, \overline{\eta}^{\overline{k}} \; U^{k - \overline{k}} \; .
\end{equation}
The sum over $k, \overline{k} \in \mathds{N}_0$ can be replaced by a sum over two new variables $p \in \mathds{Z}$ and 
$\overline{p} \in \mathds{N}_0$. They are related to the original variables $k$ and $\overline{k}$ via
\begin{equation}
k - \overline{k} \; = \; p \; \; , \; \; \; k + \overline{k} \; = \; |p| + 2 \overline{p} \; \; , \; \; \; 
k \; = \; \frac{|p| + p}{2} + \overline{p} \; \; , \; \; \; \overline{k} \; = \; \frac{|p| - p}{2} + \overline{p} \; .
\end{equation}
Using the new summation variables we find
\begin{equation}
e^{ \, - \eta \, U} \, e^{ \, - \overline\eta \, U^{-1}} \; = \; \sum_{-\infty}^\infty (-1)^p \left( \frac{\eta}{\overline{\eta}} \right)^{p/2}  U^{\,p} \; 
\sum_{\overline{p}=0}^\infty \frac{\left( \sqrt{\eta \overline{\eta}} \right)^{|p| + 2 \overline{p}} }{(|p| + \overline{p})! \; \overline{p}! } \; .
\label{app}
\end{equation}
The sum over $\overline{p}$ represents a modified Bessel function (see, e.g., \cite{nist}),
\begin{equation}
I_n(z) \; = \; \sum_{j=0}^\infty \frac{(z/2)^{2j + n}}{(n+j)! \, j!} \; ,
\end{equation}
and replacing the corresponding sum in (\ref{app}) 
by $I_{|p|} \!\left( 2 \sqrt{\eta \overline{\eta}} \, \right)$ completes the proof of (\ref{gaugebessela}).


\begin{thebibliography}{1234567}

\bibitem{reviews}
  D.~Sexty,
  PoS LATTICE {\bf 2014} (2015)  [arXiv:1410.8813 [hep-lat]].
 %
  C.~Gattringer,
  PoS LATTICE {\bf 2013} (2014) 002
  [arXiv:1401.7788 [hep-lat]].
%
  G.~Aarts,
  PoS LATTICE {\bf 2012} (2012) 017
  [arXiv:1302.3028 [hep-lat]].
%
  L.~Levkova,
  PoS LATTICE {\bf 2011} (2011) 011
  [arXiv:1201.1516 [hep-lat]].
%
  S.~Gupta,
  PoS LATTICE {\bf 2010} (2010) 007
  [arXiv:1101.0109 [hep-lat]].
%
  U.~Wolff,
  PoS LATTICE {\bf 2010} (2010) 020
  [arXiv:1009.0657 [hep-lat]].
%
  P.~de Forcrand,
  PoS LAT {\bf 2009} (2009) 010
  [arXiv:1005.0539 [hep-lat]].
%
  A.~Li,
  PoS LAT {\bf 2009} (2009) 011
  [arXiv:1002.4459 [hep-lat]].
%
  S.~Chandrasekharan,
  PoS LATTICE {\bf 2008} (2008) 003
  [arXiv:0810.2419 [hep-lat]].
%
  S.~Ejiri,
  PoS LATTICE {\bf 2008} (2008) 002
  [arXiv:0812.1534 [hep-lat]].
  
\bibitem{abelian1}
  T.~Sterling, J.~Greensite,
  Nucl.\ Phys.\ B {\bf 220} (1983) 327.
%
  M.~Panero,
  JHEP {\bf 0505} (2005) 066
  [hep-lat/0503024].
%
  M.G.~Endres,
  Phys.\ Rev.\ D {\bf 75} (2007) 065012
  [hep-lat/0610029];
%
  PoS LAT06 (2006) 133
  [hep-lat/0609037].
%
  V.~Azcoiti, E.~Follana, A.~Vaquero, G.~Di Carlo,
  JHEP {\bf 0908} (2009) 008
  [arXiv:0905.0639 [hep-lat]].
%
  T.~Korzec, U.~Wolff,
  PoS LATTICE {\bf 2010} (2010) 029
  [arXiv:1011.1359 [hep-lat]];
%
  PoS LATTICE {\bf 2013} (2014) 039
  [arXiv:1309.1331 [hep-lat]];
%
  T.~Korzec and U.~Wolff,
  Nucl.\ Phys.\ B {\bf 871} (2013) 145
  [arXiv:1212.2875 [hep-lat]].
%
  P.N.~Meisinger, M.C.~Ogilvie,
  arXiv:1306.1495 [hep-lat].
%
  C.~Gattringer, A.~Schmidt,
  Phys.\ Rev.\ D {\bf 86} (2012) 094506
  [arXiv:1208.6472 [hep-lat]].
%
  Y.D.~Mercado, C.~Gattringer and A.~Schmidt,
  Phys.\ Rev.\ Lett.\  {\bf 111} (2013) 14,  141601
  [arXiv:1307.6120 [hep-lat]];
%
  Comput.\ Phys.\ Commun.\  {\bf 184} (2013) 1535
  [arXiv:1211.3436 [hep-lat]].
%
  A.~Schmidt, P.~de Forcrand, C.~Gattringer,
  PoS Lattice 2014 (2015) 209
  [arXiv:1501.06472 [hep-lat]].
 
\bibitem{scalarqed2}
  T.~Kloiber and C.~Gattringer,
  PoS Lattice 2014 (2015) 345 [arXiv:1410.3216 [hep-lat]].

\bibitem{nucu}
  I.O.~Stamatescu,
  Phys.\ Rev.\ D {\bf 25} (1982) 1130.


\bibitem{schwinger1}
  M.~Salmhofer,
  Nucl.\ Phys.\ B {\bf 362} (1991) 641.
%
  H.~Gausterer, C.B.~Lang and M.~Salmhofer,
  Nucl.\ Phys.\ B {\bf 388} (1992) 275.
%
  H.~Gausterer and C.B.~Lang,
  Nucl.\ Phys.\ B {\bf 455} (1995) 785
  [hep-lat/9506028].
%
  F.~Karsch, E.~Meggiolaro and L.~Turko,
  Phys.\ Rev.\ D {\bf 51} (1995) 6417
  [hep-lat/9411019].
%
  K.~Scharnhorst,
  Phys.\ Rev.\ D {\bf 56} (1997) 3650
  [hep-lat/9505001].
 
\bibitem{schwinger2}
C.~Gattringer,
  Nucl.\ Phys.\ B {\bf 559} (1999) 539
  [hep-lat/9903021];
%
  Nucl.\ Phys.\ B {\bf 543} (1999) 533
  [hep-lat/9811014].

\bibitem{schwinger3}
  U.~Wenger,
  Phys.\ Rev.\ D {\bf 80} (2009) 071503
  [arXiv:0812.3565 [hep-lat]];
%
  PoS LAT {\bf 2009} (2009) 022
  [arXiv:0911.4099 [hep-lat]].
  
\bibitem{durr}
  S.~D\"urr and C.~H\"olbling,
  Phys.\ Rev.\ D {\bf 69} (2004) 034503
  [hep-lat/0311002].
 
\bibitem{shailesh}
  D.J.~Cecile, S.~Chandrasekharan,
  Phys.\ Rev.\ D {\bf 77} (2008) 054502
  [arXiv:0801.1857 [hep-lat]].

\bibitem{narayanan}  
  R.~Lohmayer and R.~Narayanan,
  Phys.\ Rev.\ D {\bf 88} (2013) 10,  105030
  [arXiv:1307.4969 [hep-th]].
  
\bibitem{nist}
  NIST Handbook of Mathematical Functions, Cambridge University Press, 2010.
      
\end{thebibliography}
\end{document}